\documentclass[12pt]{iopart}
\usepackage{graphicx}
\usepackage{url}

\begin{document}

\title[Metastable dynamics in RF micro-plasma jets]
{Spatial dynamics of helium metastables in sheath or bulk dominated rf micro-plasma jets}

\author{B~Niermann$^1$, T~Hemke$^2$, N~Y~Babaeva$^3$, M~B\"oke$^1$,
M~J~Kushner$^3$, T~Mussenbrock$^2$, and J~Winter$^1$}

\address{$^1$Ruhr-Universit\"at Bochum, Institute for Experimental
Physics II, Universit\"atsstra\ss e 150, 44801 Bochum, Germany}
\address{$^2$Ruhr-Universit\"at Bochum, Institute for Theoretical
Electrical Engineering, Universit\"atsstra\ss e 150, 44801 Bochum, Germany}
\address{$^3$University of Michigan, Electrical Engineering and Computer Science Department, 1301 Beal Ave, Ann Arbor, MI 48109-2122, USA}

\ead{benedikt.niermann@rub.de}

\begin{abstract}
Space resolved concentrations of helium He ($^3\mbox{S}_1$) metastable
atoms in an atmospheric pressure radio-frequency micro-plasma jet were
measured using tunable diode laser absorption spectroscopy. The spatial
profile of metastable atoms in the volume between the electrodes was
deduced for various electrode gap distances. Density profiles reveal
the sheath structure and reflect the plasma excitation distribution,
as well as the dominance of the $\alpha$-mode discharge. Gap width
variations show the transition from a normal glow plasma to a pure
sheath discharge. In order to analyze and verify the experimentally
observed profiles of the metastable atoms a 2-dimensional simulation
model was set up. Applying an appropriate He/N$_2$/O$_2$ chemistry model
the correlation between the metastable profiles and the underlying
excitation mechanisms was obtained.
\end{abstract}
%\submitto{\JPD}

\maketitle

\section{Introduction}

Understanding the energy transfer processes in micro-plasmas is
one of the key issues to develop new applications and reliable
process control. In this context metastable species play a
decisive role. Due to their long lifetime metastables collide
more frequently with other particles. The metastable density
in micro-discharges is several orders of magnitude lower than
the density of the ground-state atoms. However, compared to
most other species the density is significant and the electron
collision excitation cross sections of some helium levels out
of the metastable states exhibit values which are several orders
of magnitude larger and have much lower thresholds than those
for the ground state \cite{Katsch1996, Flohr1993}. Among these metastable
atoms the He ($^3\mbox{S}_1$) level plays a decisive role, since
helium is used as a feed gas in many micro-discharges. This
work reports the measurement and simulation of He ($^3\mbox{S}_1$)
atoms in an atmospheric pressure micro-plasma jet, providing a simple
design by featuring an $\alpha$-mode RF discharge between
two bare metallic electrodes.

Reliable techniques are needed for the systematic investigation
of plasma characteristics and dynamics. This is essential for
the optimization of plasma sources and process control.
Application of conventional diagnostics, especially invasive
ones, is often impossible regarding the small dimensions,
high operating pressures, and high power densities. In this
context absorption spectroscopy is a widely used technique
to measure e.g. absolute concentrations of particles in
plasma discharges, since it is non-invasive, highly sensitive
and provides a sufficient spatial resolution \cite{Niemax2005,
Tachibana2005}. We have applied tunable diode laser absorption
spectroscopy (TDLAS) to record the spectral profiles of the
lowest helium metastable state, deducing absolute densities
for various discharge conditions \cite{Niermann2010}.
The here used jet configuration is a well analyzed and value proven discharge. It was studied already under several experimental and theoretical approaches, describing the discharge characteristics in detail \cite{Schulz2007, Schaper2009, Waskoenig2010}. The new approach of the measurements at hand focuses on the variation of the gap width and determination of the sheath/bulk structure and influences on the distribution of excited species in the discharge.

To verify the experimental results and to adequately describe
the actual excitation mechanisms a reliable model is needed.
By means of the 2D fluid dynamic code \emph{nonPDPSIM} \cite{Kushner2004}
we perform simulations which take into account both the discharge dynamics between the electrodes as well as the influence of the lateral gas flow.

\section{The atmospheric pressure micro-plasma jet}

\begin{figure}
    \centering
    \includegraphics[width=0.7\textwidth]{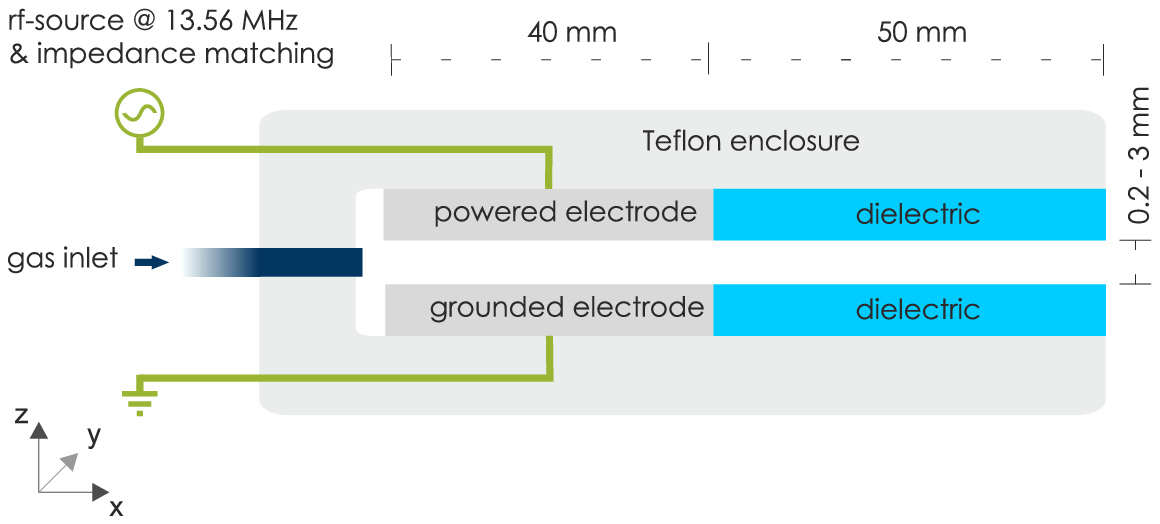}
        \caption{Sketch of the micro-plasma jet discharge, which is also the simulation domain. Shown is a 2-dimensional cross-section through plane that is spanned by the two electrodes.}
        \label{figure1}
\end{figure}

The atmospheric pressure micro-plasma jet is a capacitively coupled,
non-\-thermal glow-discharge plasma at high pressures. The design
concept of this discharge is based on the plasma jet introduced by
Selwyn et al. in 1998 \cite{Selwyn1998} and advanced by Schulz-von der
Gathen et al. \cite{Schulz2007} termed $\mu$APPJ.
The feed gas flows between two closely
spaced stainless steel electrodes driven at 13.56 MHz radio-frequency
in a parallel plate configuration (Figure \ref{figure1}). Electrodes,
plasma volume and effluent are enclosed by quartz windows, giving
direct optical access to the plasma itself and the effluent volume
behind the electrodes.
The discharge uses helium as feed gas with typical
gas velocities around 100\,ms$^{-1}$. The electric field
between the electrodes causes a breakdown in the gas
and produces a plasma with electron temperature and density of
about 1 to 2\,eV and $10^{10 }\,\mbox{cm}^{-3}$, respectively
\cite{Schaper2009, Waskoenig2010}. Atoms and molecules in the feed gas
become excited, dissociated or ionized by electron impacts. Since
the electrons are not in thermal equilibrium with the ions and
neutrals, the gas temperature remains a few tens Kelvin above
room temperature \cite{Knake2008}.

The presented jet configuration features a dielectric extension
of the gas channel to assure controlled gas flow in the effluent
behind the plasma. Electrodes and dielectric extensions are 40\,mm
and 50\,mm in length, respectively. The distance between the two
windows is fixed to 1\,mm while the electrode gap width is variable
between 0.2\,mm and 30\,mm.

\subsection{TDLAS set-up}

\begin{figure}
    \centering
    \includegraphics[width=0.7\textwidth]{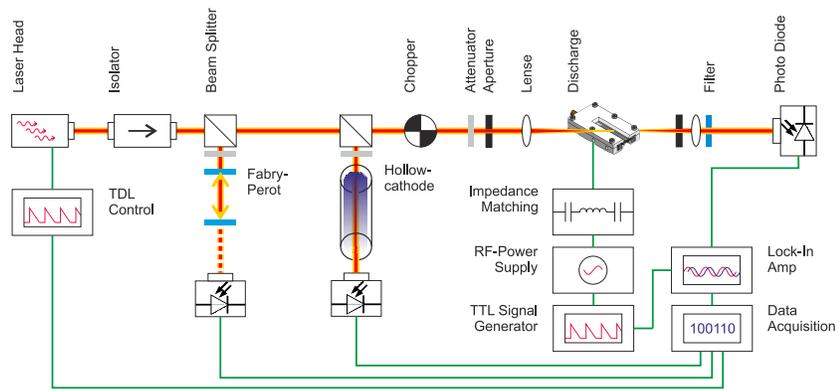}
        \caption{TDLAS setup.}
        \label{figure2}
\end{figure}

The small dimensions of micro-discharges and their operation at atmospheric pressures are a challenge for optical diagnostics, since high sensitivity and high spatial resolution are required. For the absorption spectroscopic measurements a standard TDLAS set-up was used. The absorption profile was recorded by scanning the laser frequency across the absorption line. Figure \ref{figure2} shows a sketch of the experimental setup. The laser beam from the DL passed through two beam splitters. A part of the beam was guided to a Fabry-Perot interferometer (1\,GHz free spectral range), a second part through a low pressure reference cell to perform the calibration of the laser frequency. The part of the beam transmitted through the first beam splitter was attenuated by neutral density filters with an optical density in the order of 3, and focused into the discharge with a beam power of less than 2\,$\mu$W at 100\,$\mu$m spot size, to avoid any saturation effects. After passing the discharge the beam was guided through a set of apertures and filters to suppress the emission from the plasma by reducing the collection angle and blocking wavelengths different than the observed transitions. The transmitted beam intensity was measured by photodiodes with on-chip transimpedance amplifier.
For an effective measurement of the absorption signal across the
jet axes, the discharge casing was mounted on a small movable stage
featuring three electronically controlled stepping motors to adjust,
with high precision, the discharge cell position in all spatial
dimensions. This set-up allows the positioning of the jet with an
accuracy of about 5 $\mu$m and automated $xz$-mapping of the complete
plasma volume (In $y$ direction all measurements presented in this paper
are space averaged).

Since the absorption rate of the laser light by metastable atoms is very low, in the order of $10^{-3}$ after 1\,mm absorption length, lock-in technique was used to measure the changes in signal intensity. Applying lock-in technique requires the pulsing of the signal to be measured. This was realized by pulsing the RF-power coupled into the system, which consequently leads to a pulsing of the metastable density in the discharge. The pulse frequency was chosen to 4\,kHz and a duty cycle of 50\,\%.

Absolute metastable densities were derived from the transmittance $I/I_0$ and the Beer-Lambert law. Therefore, the four signals
\begin{eqnarray*}
&L(\nu)\, &- \mbox{Plasma and Laser on},\\
&L_0(\nu)\, &- \mbox{Plasma off, Laser on},\\
&P(\nu)\, &- \mbox{Plasma on, Laser off},\\
&B(\nu)\, &- \mbox{Plasma and Laser off (background)},
\end{eqnarray*}
have to be acquired to calculate the transmittance spectra and correlate them with the plasma properties by
\begin{eqnarray}
\frac{I(\nu)}{I_0(\nu)}=\frac{L(\nu)-P(\nu)}{L_0(\nu)-B(\nu)}=e^{-k(\nu)\,l}\rm{.}
\end{eqnarray}
$I(\nu)$ and $I_0(\nu)$ are the intensities of transmitted radiation with and without the presence of absorbing species, $k(\nu)$ is the absorption coefficient and $l$ the path length through the absorbing medium \cite{Sadeghi2004}. The absorption coefficient is connected to the population density of metastable atoms by
\begin{eqnarray}
k(\nu)=\frac{1}{4\pi\varepsilon_0}\, \frac{\pi e^2}{c\, m_{\rm e} }
\, f_{ik}\, N_i \, F(\nu),
\end{eqnarray}
where $f_{ik}$ is the oscillator strength of the line, $N_i$ the density of the lower level, and $F(\nu)$ a normalized function
($\int_{0}^\infty F(\nu) d\nu=1$) representing the absorption lineshape. All other terms have their usual definitions.
The absolute metastable density can then be given by
\begin{eqnarray}
\int_0^\infty \ln\left ( \frac{I_0(\nu)}{I(\nu)}\right )d\nu=S=\frac{e^2f_{ik}l}{4\epsilon_0 m_e c}\,  N_i,
\end{eqnarray}
where $S$ is the area under the absorption curve that provides the line-averaged density of the absorbing species.

\subsection{Simulation set-up}

The 2D simulations are performed on a non-structured numerical grid
with the fluid dynamic code \emph{nonPDPSIM} which has been
originally designed and realized by Kushner and co-workers.
A detailed description of the code and a number of successful applications
can be
found in \cite{Kushner2004, Babaeva2007A, Babaeva2007B, Babaeva2009}. Basic simulations
of a radio-frequency driven He/O$_2$ plasma jet and its effluent accounting
for the plasma dynamics and the lateral gas flow can be found in \cite{Hemke2011}.
Here, we just briefly discuss the implemented equations and the underlying physics.

The code \emph{nonPDPSIM} simulates the dynamics of weakly ionized
plasmas in the regime of medium to high pressure. It takes into account
the physics and chemistry of charged particles -- electrons with mass
$m_{\rm e}$ and charge $-e$, ions with mass $m_j$ and charge $q_j$ --
and of the excited as well as the ground state neutrals (mass $m_j$).
For all species $j$, the continuity equations (particle balances)
are simultaneously solved, where $\vec{\Gamma}_j$ is the particle
flux density and $S_j$ is the source and loss term, respectively:
\begin{equation}
\label{eq:j_continuity}
\frac{\partial n_j}{\partial t} = -\nabla\!\cdot\!\vec{\Gamma}_j + S_j.
\end{equation}
The fluxes are calculated from the momentum balances in the
drift-diffusion approximation evaluated in the local center-of-mass
system. $D_j$ and $\mu_j$ are the diffusion constant and the mobility
(if applicable) of species $j$. Further, $\vec E$ is the electrical field,
and $\vec v$ is the mass-averaged advective velocity of the medium:
\begin{equation}
\label{eq:j_flux}
\vec \Gamma_j = n_j \vec{v} - D_j \nabla n_j + \frac{q_j}{|q_j|}
\mu_j n_j \vec E.
\end{equation}
For the electron fluid, additionally an energy balance equation is
solved taking into account Ohmic heating and the energy losses
due to elastic and inelastic interaction with the neutrals and ions
as well as heat conduction,
\begin{equation}
\label{eq:e_energy}
\frac{\partial}{\partial t} \left(\frac{3}{2}n_{\rm e} T_{\rm e}\right) =
\vec{j}\! \cdot\! \vec{E}
- \nabla \cdot \left( -\kappa_{\rm e}
\nabla T_{\rm e} + \frac{5}{2} T_{\rm e} \vec{\Gamma}_{\rm e} \right)
 - n_{\rm e} \sum_i \Delta \epsilon_i k_i n_i.
\end{equation}
To capture the non-Maxwellian behavior of the electrons, all
electronic transport coefficients (the mobility $\mu_{\rm e}$,
the diffusion constant $D_{\rm e}$, and the thermal conductivity
$\kappa_{\rm e}$) as well as the electronic rate coefficients in
eqs. (\ref{eq:j_continuity}) and (\ref{eq:j_flux}) are calculated
by the local mean energy method: A zero-dimensional Boltzmann equation for
the electron energy distribution $f_{\rm}(\epsilon)$ and the transport
and rate coefficients is solved for the locally applicable gas
composition and various values of the electrical field. The tabulated
data  are then consulted in dependence of the fluid dynamically
calculated electron temperature $T_{\rm e}$.

The plasma equations are coupled to a modified version of the
compressible Navier-Stokes equations which are solved for the gas density
$\rho$, the mean velocity $\vec v$, and the gas temperature $T$.
The contributions to the
energy equation from Joule heating include only ion contributions;
the heat transfer from the electrons is included as a collisional
change in the enthalpy.
The scalar pressure $p$ is given by the ideal gas law.

Finally, the potential $\Phi$ is calculated from Poisson's equation.
(The code works in the electrostatic approximation such that
$\vec E=-\nabla\Phi$.) The charge density on its right hand side stems from
the charged particles in the plasma domain and from the bound charges
$\rho_s$ at the surfaces. The coefficient $\varepsilon=\varepsilon_0
\varepsilon_{\rm r}$ represents the permittivity of the medium:
\begin{equation}
-\nabla \cdot \varepsilon \nabla \Phi = \sum_j q_j n_j + \rho_{\rm s}.
\end{equation}
The surface charges are governed by a separate balance equation,
where $\sigma$ is the conductivity of the solid materials and the
subscript $s$ indicates evaluation on the surface:
\begin{equation}
\frac{\partial \rho_{\rm s}}{\partial t} = \left[ \sum_j q_j
(-\nabla \cdot \vec{\Gamma}_j + S_j) -
\nabla \cdot \left( \sigma (-\nabla \Phi) \right) \right]_{\rm s}.
\end{equation}

The dynamical equations are complemented by an appropriate set of
boundary conditions. Electrically, the walls are either powered or
grounded. With respect to the particle flow, they are either solid,
or represent inlets or outlets: The  flow is specified to a given
flux, while the outlet flow is adjusted to maintain the pressure.
Finally, it is worth mentioning that the actual implementation of
the equations poses some difficulties due to the vast differences
in the time scales of the dynamics of the plasma and the neutrals.
These difficulties are overcome by the methods of time-slicing
and subcycling.

The described code is employed to simulate the $\mu$APPJ depicted in Figure \ref{figure1}
with varying electrode gap sizes of 1.8\,mm, 1.0\,mm and 0.3\,mm. We choose these
values to characterize the three distinct regimes of plasma bulk-sheath ratios.
The simulation resolves the two Cartesian dimensions $x$ and $z$. In direction $y$, translational invariance is assumed.

By leaks in the gas supply system as well as intrusion of air
from the exit nozzle of the jet,
the amount of nitrogen and oxygen in the feed gas is significant and has to be taken
into account for the modeling approach.
We mimic the impurities of the experimental setting according to \cite{Niermann2011}
by adding 0.016 \% of nitrogen and 0.004 \% oxygen in the chemical model
(Oxygen species contribute to the electronegativity of the plasma while nitrogen has an impact on the quenching of the helium metastable density.).
We consider the following species:
Ground state neutrals O$_2$, O, O$_3$, and He,
O$_2(\nu)$, representing the first four vibrational levels of
O$_2$, the electronically excited states O$_2(^1\Delta_g)$,
O$_2(^1\Sigma^+_g)$, O$(^1D)$, O$(^1S)$, and He$^*$ $\equiv$ He$(^3 \rm S_1)$,
He$_2^*$ $\equiv$ He$_2(^3 \Sigma^+_u)$,
positive ions
O$_2^+$, O$^+$, He$^+$ and He$_2^+$,
negative ions O$_2^-$, O$^-$, and O$_3^-$, and electrons.
The flow rate of the gas mixture injected in the jet is adjusted to
achieve advective gas velocities of about 100\,ms$^{-1}$.
The outlet  is controlled to maintain a constant pressure. \linebreak
Finally, the secondary electron emission coefficient of
the electrode surfaces depends on the ion species and varies from
$0.26$ for He$^+$ to $0.06$ for O$_2^+$.

\section{Results and discussion}

\subsection{Spatial distribution of He metastable atoms}

\begin{figure}
    \centering
    \includegraphics[width=0.7\textwidth]{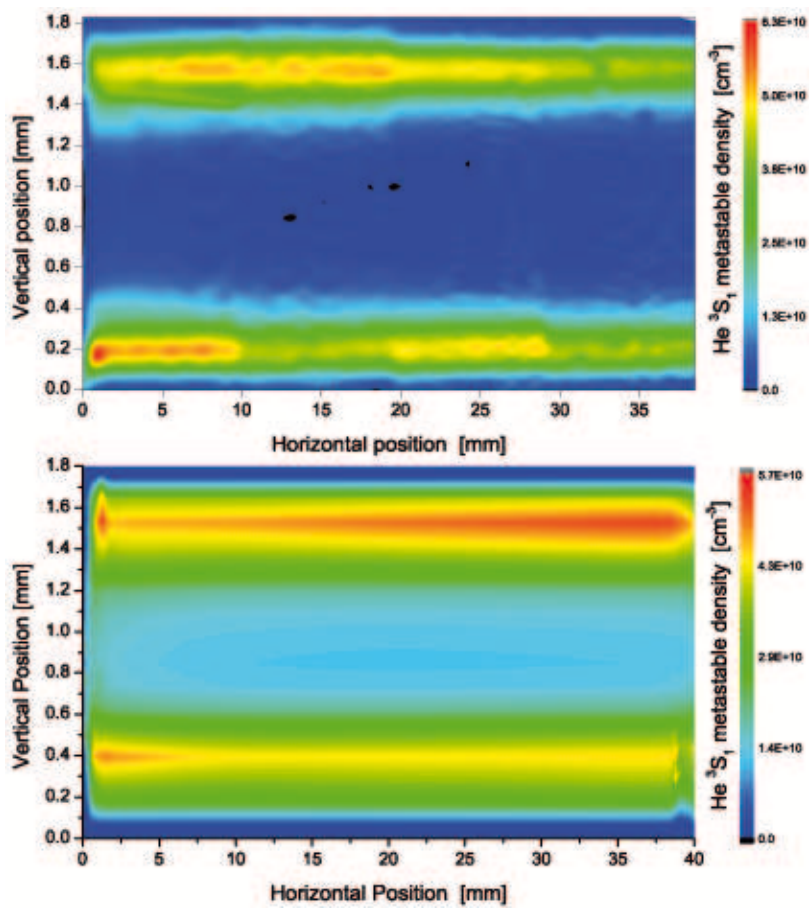}
        \caption{\textbf{Top:} 2-dimensional map of the measured metastable density in the discharge volume for the He $2^3\mbox{S}_1\rightarrow\,2^3\mbox{P}^0_{1,2}$ transition. Powered electrode at the top, grounded electrode at the bottom. Electrode gap width was 1.8\,mm. Densities are given in cm$^{-3}$. \textbf{Bottom:} Simulation results for the spatial distribution of the He metastables.}
        \label{figure3}
\end{figure}

Figure \ref{figure3} shows 2D-maps of the He ($^3\mbox{S}_1$) metastable
density in the discharge volume. The upper map shows experimental results
whereas the bottom map shows simulation data. The horizontal and vertical
axis span the exact area between the electrodes. The map of the experimental
data, which covers 2,000 reading points (40 vertical x 50 horizontal) of the
absorption signal, in the plasma volume is in good qualitative and quantitative
agreement with the simulation results. It should be mentioned that the steps occurring in the measured map at certain positions are due to mechanical limitations in the setup, that made it necessary to record that map in four successive intervals. Furthermore the measurement time for the map was several days, explaining why the map shows heavy fluctuations, whereas the actual metastable distribution in the discharge is more smooth, as seen in the measurements presented later in this article.

A variety of effects determines the metastable distribution in the horizontal
and the vertical axis. In horizontal direction the density profile is
governed mainly by two effects. One is the alignment of the electrodes,
since the electric field distribution strongly influences the electron energy.
In the present experimental case the electrode gap is marginally larger on the
rear side of the jet, resulting in a slight decrease of densities from
left to right. A second impact is given by impurities entering the jet
through the exit nozzle, therewith quenching the metastable atoms. Since
the jet is running permanently in contact with the ambient atmosphere the
intrusion of nitrogen and oxygen into the plasma channel is significant,
and decreases the metastable density especially in the first millimeters
from the nozzle. The small fluctuations in the horizontal profile (which are not
resolved by the simulation model) can be attributed to turbulences, imperfections
in the surface of the electrodes or singularities in the walls of the gas channel.

\begin{figure}
    \centering
    \includegraphics[width=0.7\textwidth]{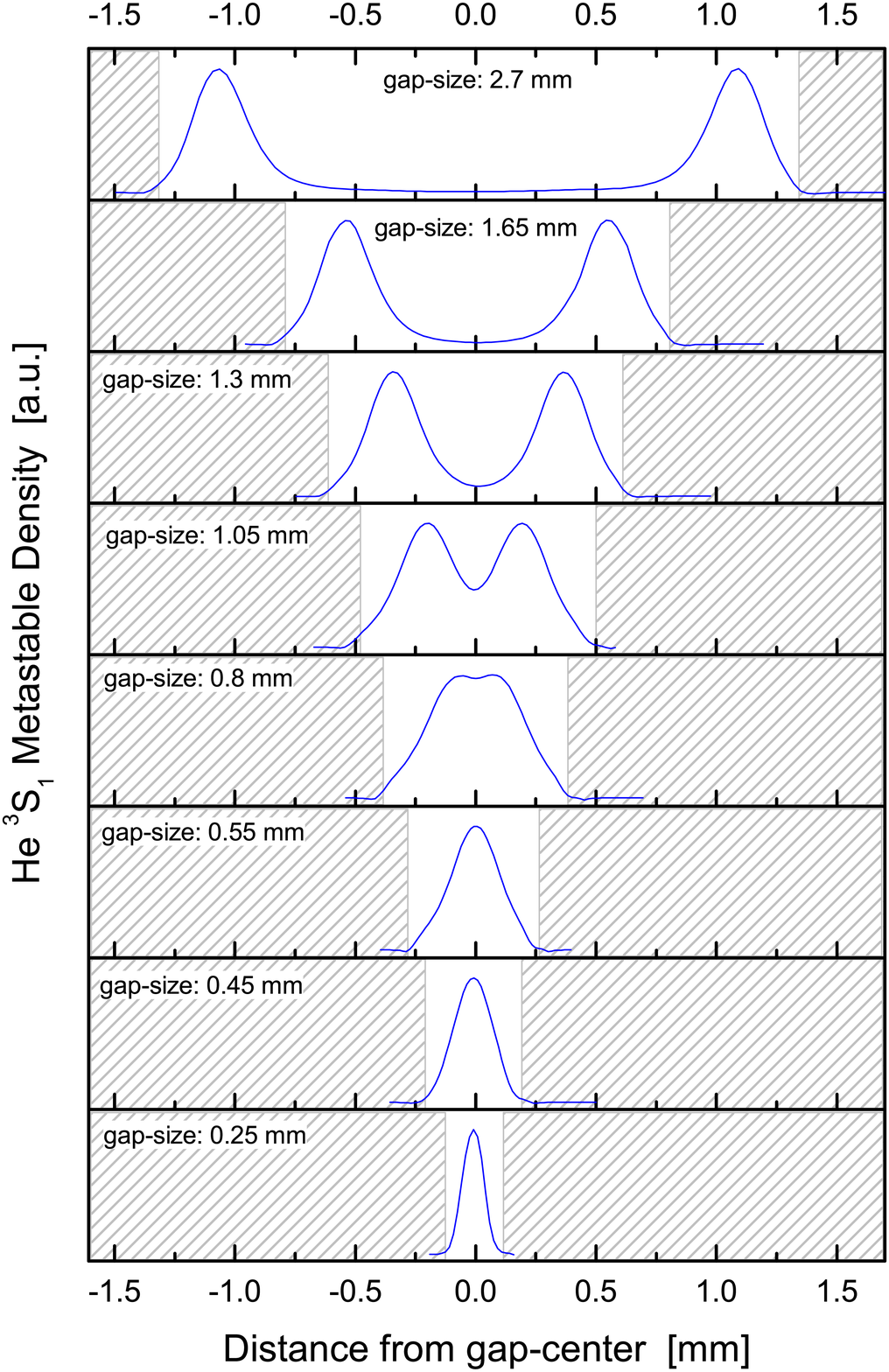}
        \caption{Measured vertical metastable profile for gap widths between 2.7\,mm and 0.25\,mm.}
        \label{figure4}
\end{figure}

The vertical metastable profile is less determined by these environmental
factors but by the electron density and temperature distribution. The
observed structure is consistent with the sheath structure in an RF discharge.
Directly in front of the electrodes, the metastable density is low since
the electron density is too low for an efficient excitation and ionization
of the ground state atoms (see also Figure \ref{figure4}, below). Metastables reach
their highest densities some 10\,$\mu$m away from the electrode at the
plasma-sheath interface, where in the negative glow area most excitation
and ionization processes occur. Here in the sheath/bulk interface the electron
energy is highest. In the bulk, the electric field strength and therewith the
electron temperature is low (see also Figure \ref{figure5}, below), yielding to a lower
metastable production rate. The asymmetry in the vertical profile of the simulated data can be attributed to stochastic effects in the simulation since it is not averaged over many RF-cycles due to runtime considerations, while the measurements are averaged over several 10$^{6}$ cycles.

\subsection{Gap width variations}

Due to the variability of the electrode gap the micro-plasma jet offers the
possibility to continuously tune the sheath/bulk ratio and therewith the
vertical excitation profile. Figure \ref{figure4} shows the He metastable profile for
gap widths between 2.7 and 0.25\,mm. These values represent the limits where
the pure helium discharge can be operated in $\alpha$-mode. With decreasing gap width the
breakdown voltage decreases while approaching the Paschen minimum. Although the outer power
coupling to the discharge for each gap width is different, one can clearly observe a systematic
increase of the metastable density in the maximum with smaller gap width. A quantitative description of this increase is difficult, since the gap width correlates with a variety of discharge parameters, like the impedance, the power coupling or the gas velocity, that all influence the species' density.

\begin{figure}
    \centering
    \includegraphics[width=0.7\textwidth]{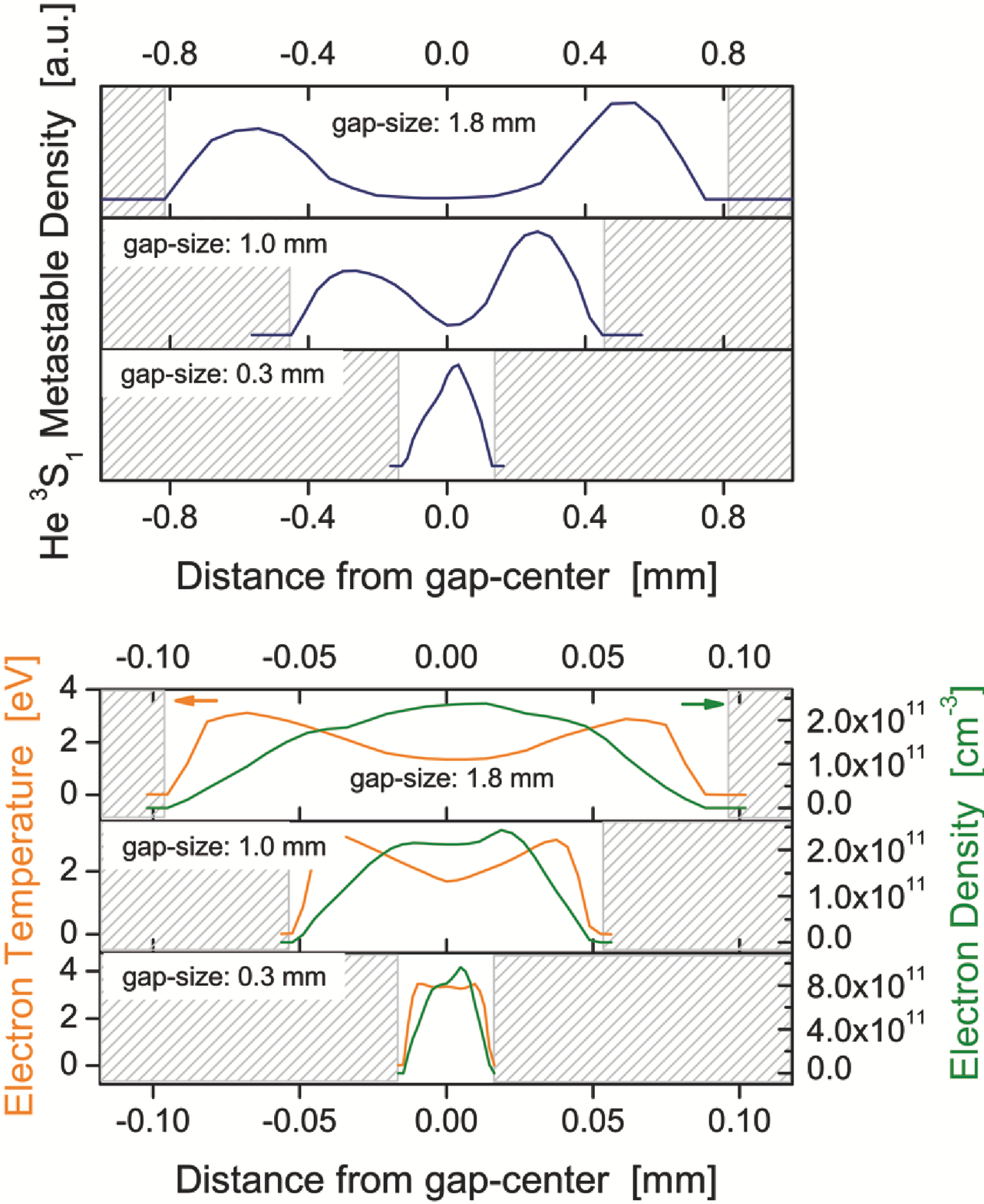}
        \caption{\textbf{Top:} Simulation results of characteristic metastable profiles for various gap widths. \textbf{Bottom:} Simulation results of  electron density and electron temperature profiles for various gap widths.}
        \label{figure5}
\end{figure}

The graphs show a steep rise of the density in front of the electrode surface with a maximum
in the sheath/bulk interface, while in the bulk the density is low. The
density distribution suggests that diffusion processes play a minor role for
the metastable profile. At atmospheric pressure the mean free path of the
electrons is short. High energetic electrons are lost in the sheath
and can not contribute to the excitation in the plasma bulk. Although the
metastable lifetime is comparatively long, in the order of 1\,$\mu$s, the
metastable diffusion to the bulk can be neglected likewise \cite{Niermann2011}. De-excitation
of these species is dominated by three-body collision with ground state
atoms as well as quenching reaction with impurities, like molecular nitrogen
and oxygen. A detailed analysis of metastable lifetimes in this discharge is given in \cite{Niermann2011}.
For gap widths larger than 1\,mm the two excitation areas are
clearly separated by the plasma bulk. Lowering the electrodes distance
leads to the coalescence of the two sheath regions, and an excitation
profile where highest densities are in the center of the gap. For electrode
distance of just a few hundred $\mu$m the plasma transforms to a pure
``sheath discharge''. These experimental findings are supported by the results
from the numerical simulation. Figure \ref{figure5} (top) shows the metastable profile
for selected gap widths, the related electron densities and electron temperatures
are given in Figure \ref{figure5} (bottom). Due to the admixture of oxygen into the feed gas the electronegativity of the plasma has to be taken into account. The electron temperature profile for a gap widths of 1.8\,mm represents the excitation pattern of the metastable atoms. The temperatures are highest in the sheath/bulk interface, with a maximum of about 3\,eV. Lowering the gap widths results, like in the experiment, in a convergence of the two sheaths. For a low gap widths of 0.3\,mm the electron density increases significantly, while temperatures reach values of almost 4\,eV. This behavior suggests that the high energy tail of the EEDF is getting more pronounced. Since the high electron energies enhance the production of metastable atoms, the simulation again supports the qualitative observation of very high metastable densities for low gap widths.\\

For metastable density measurements no error bars are shown, since statistical as well as systematic errors are negligible. For the presented data they are in the order of $10^{9}\,\rm{cm}^{-3}$, while the density values are in the order of $10^{11}\,\rm{cm}^{-3}$ to $10^{13}\,\rm{cm}^{-3}$. Due to the use of lock-in technique a sensitive measurement of low densities was possible with a high signal to noise ratio. A minor systematic error is caused by the unknown metastable profile in y-axis, and the implications on the absorption length. But since the metastable production and loss processes take place on a very local scale at atmospheric pressure, the profile in y-axis is expected to be homogeneous with the exception of a tiny volume in front of the glass windows that confines the plasma channel and acts as a sink for metastable atoms.

\section{Summary}

We have shown that the experimental results for metastable distributions in
micro-plasma jet discharges can be reproduced qualitatively an quantitatively
by a 2D numerical model, therewith revealing the underlying excitation mechanisms
of the discharge. By varying the gap width it was proven that the simulation model
reliably mimic the experimental results for a large span of sheath bulk ratios.
This offers now new possibilities in the analysis of extreme discharge conditions,
like $\gamma$-mode structures and very small gap widths showing effects like no
quasi-neutrality all over the discharge volume.

\clearpage

\ack
This project is supported by DFG (German Research Foundation) within the framework of the Research Unit FOR 1123 and the Research Department 'Plasmas with Complex Interactions' at Ruhr-University Bochum.

%%%%%%%%%%
%%%%%%%%%%

\clearpage

\clearpage

\clearpage

\clearpage

\clearpage

\clearpage

\section*{References}

\begin{thebibliography}{20}

\bibitem{Katsch1996}{Katsch H M, Quand E and Schneider T 1996
{\it Plasma Physics and Controlled Fusion} {\bf 38} 183}

\bibitem{Flohr1993}{Flohr R, Melzer A and Piel A 1993
{\it Plasma Sources Sci. Technol.} {\bf 3} 206}

\bibitem{Niemax2005}{Miclea M, Kunze K, Heitmann U, Florek S, Franzke J and Niemax K 2005
{\it J. Phys. D, Appl. Phys.} {\bf 38} 1709}

\bibitem{Tachibana2005}{Tachibana K, Kishimoto Y and Sakai O 2005
{\it Journal of Applied Physics} {\bf 97} 123301}

\bibitem{Niermann2010}{Niermann B, B\"oke M, Sadeghi N and Winter J 2010
{\it Eur. Phys. J. D} {\bf 60} 489}

\bibitem{Schulz2007}{Schulz-von der Gathen V, Buck V, Gans T, Knake N, Niemi K, Reuter S,
Schaper L and Winter J 2007
{\it Contrib. Plasma Phys. }{\bf 47} No. 7 510}

\bibitem{Schaper2009}{Schaper L, Reuter S, Waskoenig J, Niemi K, Schulz-von der Gathen V and
Gans T 2009
{\it  J. Phys. Conf.} {\bf 162} 012013}

\bibitem{Waskoenig2010}{Waskoenig J, Niemi K, Knake N, Graham L M, Reuter S,
Schulz-von der Gathen V and Gans T 2010
{\it  Plasma Sources Sci. Technol.} {\bf 19} 045018}

\bibitem{Kushner2004} Kushner M J 2004
{\it J. Appl. Phys.} \textbf{95} 846

\bibitem{Selwyn1998}{Sch\"utze A, Jeong J Y, Babayan S E, Park J, Selwyn G S
and Hicks R F 1998
{\it  IEEE Trans. Plasma Sci.} {\bf 26} 1685}

\bibitem{Knake2008}{Knake N, Reuter S, Niemi K, Schulz-von der Gathen V
and Winter J\"org  2008
{\it  J. Phys. D: Appl. Phys.} {\bf 41} 194006}

\bibitem{Sadeghi2004}{Sadeghi N 2004
{\it  J. Plasma Fusion Res.} {\bf 80} 767}

\bibitem{Babaeva2007A}{Babaeva N Y, Arakoni R and Kushner M J 2007
{\it J. Appl. Phys.} {\bf 101} 123306}

\bibitem{Babaeva2007B} Babaeva N Y and Kushner M J 2007
{\it J. Appl. Phys.} \textbf{101} 113307 (2007)

\bibitem{Babaeva2009} Babaeva  N Y and Kushner M J 2009
{\it Plasma Sources Sci. Technol} \textbf{18} 035009

\bibitem{Hemke2011} Hemke T, Wollny A, Gebhardt M, Brinkmann R P and Mussenbrock T 2011
{\it  J. Phys. D: Appl. Phys.} {\bf 44} 285206

\bibitem{Niermann2011} Niermann B, Kanitz A, B\"oke M and Winter J 2011 {\it  J. Phys. D: Appl. Phys.} {\bf 44} 325201

\end{thebibliography}
\end{document}